\documentclass[10pt]{iopart}
\input{iopams.sty}
\usepackage{graphicx}
\usepackage[dvips]{color}

\newcommand{\be}{\begin{equation}}

\newcommand{\ee}{\end{equation}}
\newcommand{\bea}{\begin{eqnarray}}
\newcommand{\eea}{\end{eqnarray}}

\begin{document}
\article[Charge Fluctuations along the QCD phase boundary]{}{Charge Fluctuations along
the QCD phase boundary}
\date{\today}
\author{ K. Redlich$^{a}$,  B. Friman$^b$, and C. Sasaki$^b$}
\address{$^a$ Institute of Theoretical Physics, University of Wroclaw, PL--50204
Wroc\l aw, Poland}
\address{$^b$ Gesellschaft f\"ur Schwerionenforschung, GSI,  D-64291 Darmstadt,
Germany}

\begin{abstract}
 We discuss  the properties  of the  net--quark  and isovector
fluctuations  along the chiral phase transition line  in the plane spanned by
temperature and baryon chemical potential. Our  results  are obtained  in terms of the
Nambu--Jona-Lasinio (NJL) model within the mean-field approximation. The model is
formulated at the finite temperature and for non-vanishing net--quark and the isospin
chemical potentials. The fermion interactions are controlled by the strength of the
scalar and vector couplings in the iso-scalar and iso-vector channels of constituent
quarks. We explore properties and differences in the behavior of the net--quark number
and isovector susceptibilities for different values of thermal parameters near the phase
transition. We argue that any non-monotonic behavior of the net--quark number
susceptibility  along the phase transition boundary  is an excellent probe of the
existence and the position of the second order endpoint in the QCD phase diagram.
\end{abstract}

One of the essential predictions of  QCD is the existence of the  boundary line in the
temperature and net--quark chemical potential, $(T,\mu_q)$--plane that separates the
confined, chirally broken hadronic phase from the deconfined quark--gluon plasma phase.
The existence of such a boundary line for $\mu_q/T\leq 1$ has been recently established
by the first principle calculations in the Lattice Gauge Theory (LGT) formulated at
finite baryon density
 \cite{fodor,lgt1,lgtm,lgtp}.

Arguments based on effective models
\cite{ef1,Klevansky,ef2,ef3,ef4,ef5,ef6,ef7,Alford:2003fq} indicate that at large $\mu_q$
the transition along the boundary line is the first order. For small $\mu_q$ and for two
massless flavor QCD the chiral transition was argued \cite{pisarski} to be second order
with the critical exponents of the O(4) spin model. For the finite quark mass, due to the
explicit chiral symmetry breaking,   this transition is likely to be replaced by the
rapid crossover. Such a different nature of the phase transition at low and high $\mu_q$
suggests that the QCD phase diagram should exhibit a critical endpoint at which the line
of the first order phase transition matches that of the second order or analytical
crossover \cite{stephanov}. The critical properties of this second order chiral endpoint
are expected to be determined by the three-dimensional Ising model universality class \cite{ef5,ising}.

The existence of  a critical endpoint in QCD has been recently studied in the lattice
calculations at the non-vanishing chemical potential by either considering the location
of Lee--Yang zeros in (2+1)--flavor QCD  \cite{fodor,shinji} or by analyzing the
convergence radius of the Taylor series in $\mu_q/T$ expansion of  the free energy
\cite{lgt1,lattice:ejiri,gupta}. Recent results \cite{fodor} based on the first approach
suggest that a critical endpoint indeed exists in QCD phase diagram and might occur at
$T\simeq 164$ and  $\mu_q\simeq 120$ MeV. In the 2--flavor QCD  and with a relatively
large quark mass  used in the actual lattice calculations  \cite{lattice:ejiri}  no
direct evidence for the existence of the critical endpoint has been found for $\mu_q<T$
where the Taylor expansion method is applicable.

The critical behavior and the position of the  chiral endpoint can be possibly identified
by observables that are sensitive to singular part of the free energy \cite{rg3}. One
such observable is the quark susceptibility $\chi_{ij}$ defined as the second order
derivative of the thermodynamical-potential $\Omega(T,\vec\mu ,V)$ with respect to
quark--flavor   chemical potentials, $ \chi_{ff^\prime} = - {\partial^2
\Omega}/{\partial\mu_f
\partial\mu_{f^\prime}}
$, where for two light (u,d)--quarks,  $\vec\mu =(\mu_u,\mu_d)$.
 To identify the chiral critical endpoint in the QCD phase
diagram,  the properties of  the net--quark number
susceptibility $\chi_q$ are of particular interest~\cite{hatta,Fujii:2003bz,Fujii:2004jt}.
In two--flavor QCD,
$\chi_q$ is expressed as the  sum of $u$ and $d$ quarks susceptibilities:
$\chi_q=2(\chi_{uu}+\chi_{ud})$. The universality argument predicts that independent of
the values of the quark masses  the $\chi_q$ should diverge at the chiral endpoint.

The   quark susceptibilities have been recently obtained
\cite{lgtm,lattice:ejiri,gupta,lgtold} within  the lattice QCD  for two light quark
flavors using  p4--improved staggered fermion action with the quark mass $m_q/T=0.4$
\cite{lgt1,lattice:ejiri}. The Monte Carlo simulations have been done at the  finite
quark chemical potential via Taylor series expansion. The susceptibilities were
calculated up to the $O(\mu_q^4)$ order  in the quark chemical potential
\cite{lattice:ejiri,rg2,rg3}. The lattice results \cite{lattice:ejiri,rg2}  for the
net--quark number $\chi_q$ and isovector susceptibility $\chi_I$  as a function of the
temperature for different values of $\mu_q$  are shown in Fig. 1.  There is a strong
suppression of the $\chi_q$ fluctuations in confined phase and the cusp-like structure in
the near vicinity of the transition temperature $T_0$. A rather strong increase  of
$\chi_q$ at the transition temperature $T_0$ with increasing quark chemical potential is
also observed on the lattice. The lattice results confirmed that the quark fluctuation in
the isovector channel $\chi_I$ contrary to $\chi_q$ does not exhibit a peak structure at
$T_0$ and shows a rather weak dependence on $\mu_q$ as seen in Fig. 1--right. Such a
properties of $\chi_q$ and $\chi_I$ are expected when approaching the chiral endpoint
with increasing $\mu_q$. However, the above behavior of $\chi_q$ and $\chi_I$ can be also
quantified by the regular part of the free energy due to the enhanced contribution of
resonances in the near vicinity of the transition temperature
\cite{lattice:ejiri,rg3,rg2,rg1}. The above became clear in Fig. 1 where the LGT results
on $T$ and $\mu_q$ dependence of different quark susceptibilities in confined phase   are
seen to be quite successfully described by the resonance gas partition function
\cite{rg2}. The apparent agreement of the LGT results with the  hadron resonance gas on
the level of the equation of state and the susceptibilities in the temperature range
$T<T_0$ indicate that increase of the $\chi_q$ fluctuations  observed in Fig. 1
 with increasing $\mu_q$ at $T_0$ is  a necessary but not
sufficient condition to verify the existence of the chiral endpoint. In the following we
will argue in model calculations that to verify the appearance of the TCP in the QCD
phase diagram would require a non-monotonic behavior of the $\chi_q$ when going along the
phase boundary.

\begin{figure}
\begin{center}
\includegraphics[width=7.4cm]{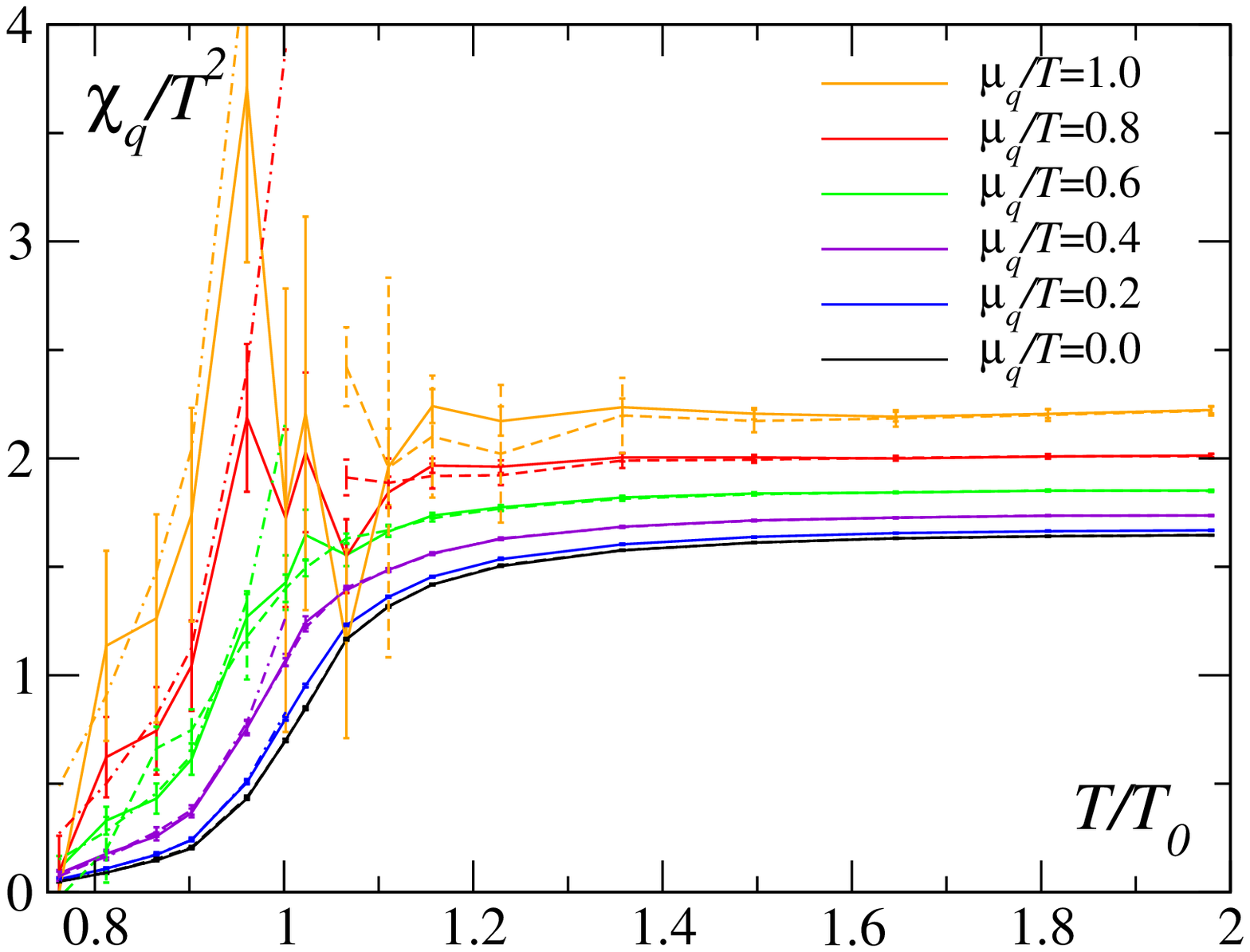}
\hskip 0.4cm \includegraphics[width=7.4cm]{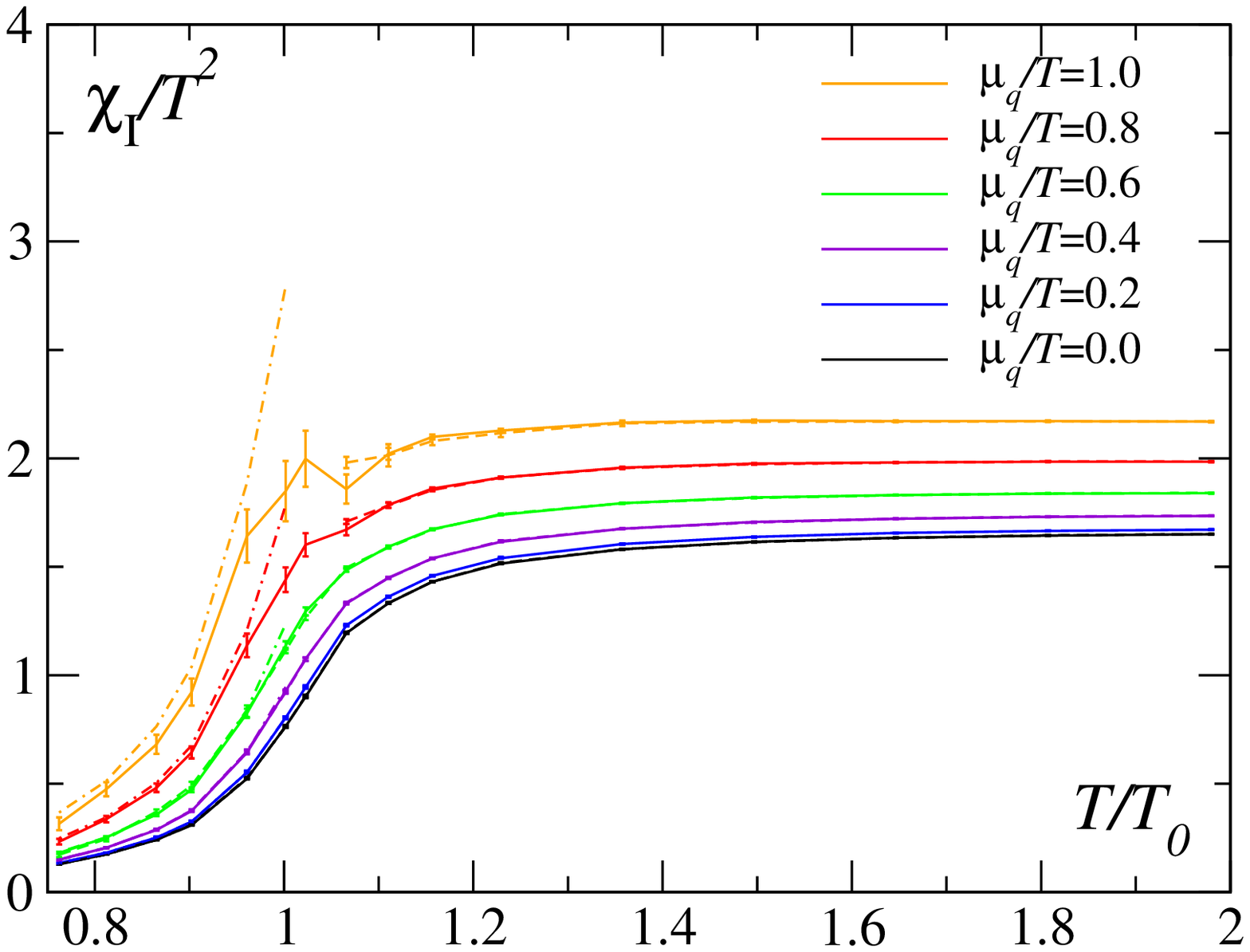} \caption{\label{fig:Mu} The left-hand
figure shows the LGT results obtained in  2--flavor QCD \protect\cite{lattice:ejiri,rg2}
for net--quark number susceptibility $\chi_q$ as a function of the temperature normalized
to $T_0$ being the transition temperature at $\mu_q=0$ . The results are shown for
different values of $\mu_q/T$. The dashed--dotted lines indicate the corresponding hadron
resonance gas model results \protect\cite{rg2}.
   The
right-hand figure: as in the left-hand figure but for the isovector susceptibility.
 }
\end{center}
\vspace*{0.5cm}
\end{figure}

\begin{figure}
\begin{center}
\includegraphics[width=9.0cm,height=7.0cm]{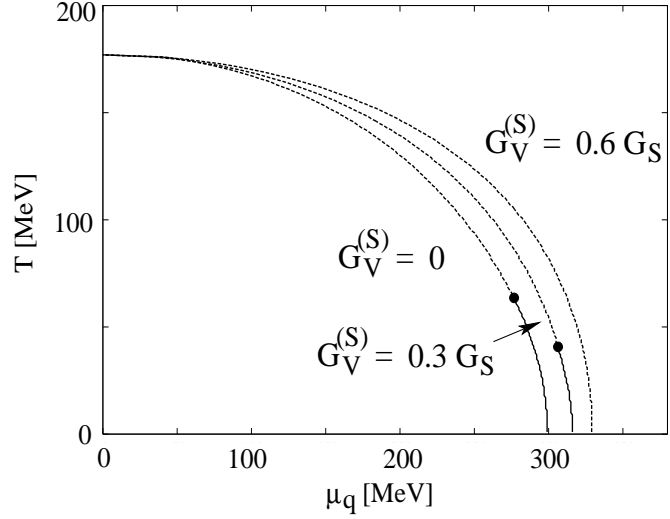}
\caption{\label{fig:phase} The NJL model phase diagram in the chiral limit for $G_V^{\rm
(S)} = 0, 0.3$ and $0.6\,G_S$. The dashed (solid) line denotes the second-order
(first-order) transition line. The tricritical point indicated by a dot ($\bullet$) is
located at $(T,\mu_q)=(65, 275)$ MeV for $G_V^{\rm (S)}=0$ and $(T,\mu_q)=(42, 305)$ MeV
for $G_V^{\rm (S)}=0.3\,G_S$. The results  are shown for vanishing isovector chemical
potential.}
\end{center}
\vspace*{0.5cm}
\end{figure}

Having in mind the importance of  the quark number fluctuations as a probe of phase
structure of QCD as well as  the above   lattice results the main scope of this article
is to  consider the properties of different quark susceptibilities in terms of an
effective chiral model. Of particular interest is the  behavior of the quark number
correlations in different channels along the boundary line and in  the near vicinity of
the chiral endpoint. The calculations will be done in terms  of two-flavor
Nambu--Jona-Lasinio (NJL) model~\cite{Nambu} formulated at the finite temperature and
chemical potentials related with baryon number and isospin conservation.

\section{The net--quark and isovector fluctuations in the NJL model}

The thermodynamics of the NJL model at the finite temperature and non vanishing
net--quark and isospin    chemical potentials is obtained from the partition function
$Z(T,\mu_q,\mu_I,V)$ formulated as a generating functional in the Euclidean space. In the
mean field approximation \cite{Klevansky} the partition function is obtained from the
effective Lagrangian

\begin{eqnarray}
{ L} = \bar{\psi}(i{\partial} - M + \tilde{\mu}\gamma_0 )\psi
 {}- \frac{1}{4G_S}(M - m)^2
{}+ \frac{1}{4G_V^{\rm (S)}}(\tilde{\mu}_q - \mu_q)^2
 {}+ \frac{1}{4G_V^{\rm (V)}}(\tilde{\mu}_I -
 \mu_I)^2\,,\label{eq2.4}
\end{eqnarray}
in which   the  thermal averages  $\langle \bar{\psi}\tau^{1,2}\gamma_i \psi \rangle$
have been  neglected. The strength of the constituent quarks interactions in scalar and
vector channels are parameterized by  the effective coupling $G_S$ and $G_V$ respectively
which  have dimensions of length square. To distinguish quark-antiquark interactions in
the  iso-scalar and iso-vector channels  we have defined two independent
couplings $G_V^{\rm (S)}$  and $G_V^{\rm (V)}$ respectively. In Eq. (\ref{eq2.4}) we have
also introduced  a dynamical mass  $M$ and
 shifted chemical potential $\tilde{\mu}$ which  are obtained
from
\begin{eqnarray}
M = m - 2G_S\langle \bar{\psi}\psi \rangle,~~~~ \tilde{\mu} = \tilde{\mu}_q +
\tilde{\mu}_I\tau^3\,,\label{eq2.5}
\end{eqnarray}
with
\begin{eqnarray}
\tilde{\mu}_q = \mu_q - 2G_V^{\rm (S)}\langle \bar{\psi}\gamma_0\psi \rangle ,~~~~
\tilde{\mu}_I = \mu_I - 2G_V^{\rm (V)}\langle \bar{\psi}\tau^3\gamma_0\psi\rangle\,.
\label{muq-muI}\label{eq2.6}
\end{eqnarray}
In the mean field approximation the thermodynamic potential of the NJL model   is
obtained in the following form
\begin{eqnarray}
\Omega (T,\mu;M,\tilde{\mu}) = \sum_{f=u,d}\Omega_f(T,\mu;M_f,\tilde{\mu}_f)  +
\nonumber\\
\frac{1}{4G_S}(M - m)^2 {}- \frac{1}{4G_V^{\rm (S)}}(\tilde{\mu}_q - \mu_q)^2 {}-
\frac{1}{4G_V^{\rm (V)}}(\tilde{\mu}_I - \mu_I)^2\,,\label{eq2.7}
\end{eqnarray}
with
\begin{eqnarray}
\Omega_f (T,\mu;M_f,\tilde{\mu}_f)=- 2 N_c \int\frac{d^3p}{(2\pi)^3}
 \Bigl[
  E_f - T\ln ( 1-n_f^{(+)}(T,\tilde{\mu}_f) )
  {}- T\ln ( 1-n_f^{(-)}(T,\tilde{\mu}_f) )
 \Bigr]\,,
\label{omega f}
\end{eqnarray}
where $E_f = \sqrt{|\vec{p}|^2 + M_f^2}$ is a quasiparticle energy and
$n_f^{(\pm)}(T,\tilde{\mu}_f) = \Bigl( 1 +
  \exp\bigl[ (E_f \mp \tilde{\mu}_f)/T \bigr] \Bigr)^{-1}$
 is the
distribution function for the particle $(+)$ and anti-particle $(-)$.

The condensates appearing in Eqs. (\ref{eq2.5})-(\ref{eq2.7}) are obtained from the
conditions to minimize  the  thermodynamic potential with respect to the dynamical mass
and the shifted chemical potentials, ${\partial\Omega}/{\partial M}
 = {\partial\Omega}/{\partial\tilde{\mu}_q}
 = {\partial\Omega}/{\partial\tilde{\mu}_I} = 0$. From the above  stationary conditions   and from Eqs.
(\ref{eq2.5})-(\ref{eq2.7}) one calculates  the quark condensates as the solution of the
following   gap equations:

\begin{eqnarray}
M_f &= m_f + 4G_S N_c \sum_{f=u,d}\int\frac{d^3 p}{(2\pi)^3} \frac{M_f}{E_f} \Bigl[ 1 -
n_f^{(+)}(T,\tilde{\mu}_f) - n_f^{(-)}(T,\tilde{\mu}_f) \Bigr]\,, \label{gap eq-mass}
\\
\mu_q &= \tilde{\mu}_q + 4G_V^{\rm (S)} N_c \sum_{f=u,d} \int\frac{d^3 p}{(2\pi)^3}
\Bigl[ n_f^{(+)}(T,\tilde{\mu}_f) - n_f^{(-)}(T,\tilde{\mu}_f) \Bigr]\,, \label{muq}
\\
\mu_I &= \tilde{\mu}_I + 4G_V^{\rm (V)} N_c \int\frac{d^3 p}{(2\pi)^3} \Bigl[ \Bigl(
n_u^{(+)}(T,\tilde{\mu}_u) - n_u^{(-)}(T,\tilde{\mu}_u) \Bigr) {}- ( u \to d ) \Bigr]\,.
\label{muI}
\end{eqnarray}
The above gap equations together with the potential (\ref{eq2.7}) are sufficient to
describe within the  NJL model the thermodynamics and the phase structure of an effective
quark medium at the finite temperature and  the  net--quark and  isovector chemical
potentials.

\begin{figure}
\begin{center}
\includegraphics[width=7.4cm]{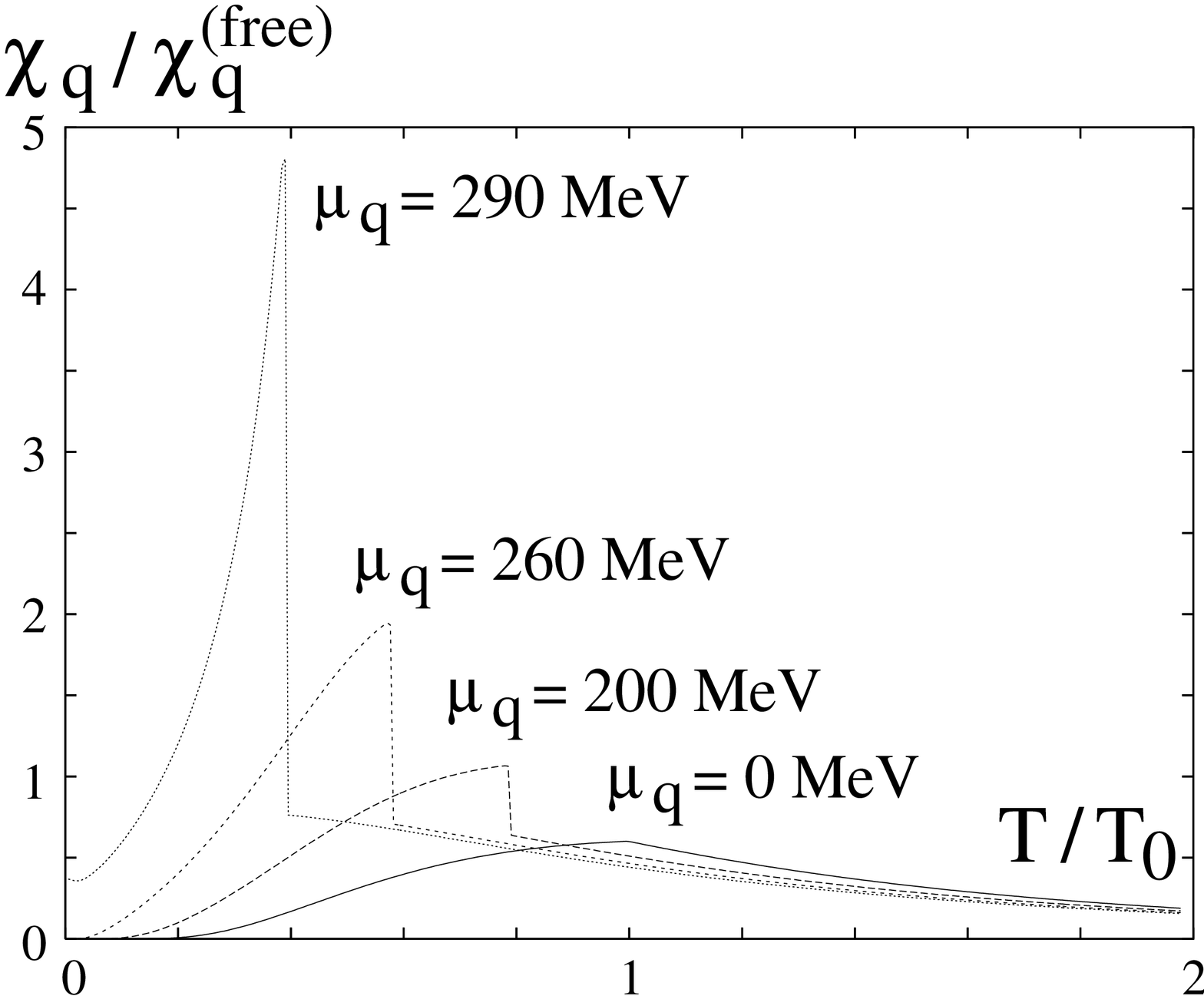}
\includegraphics[width=7.4cm]{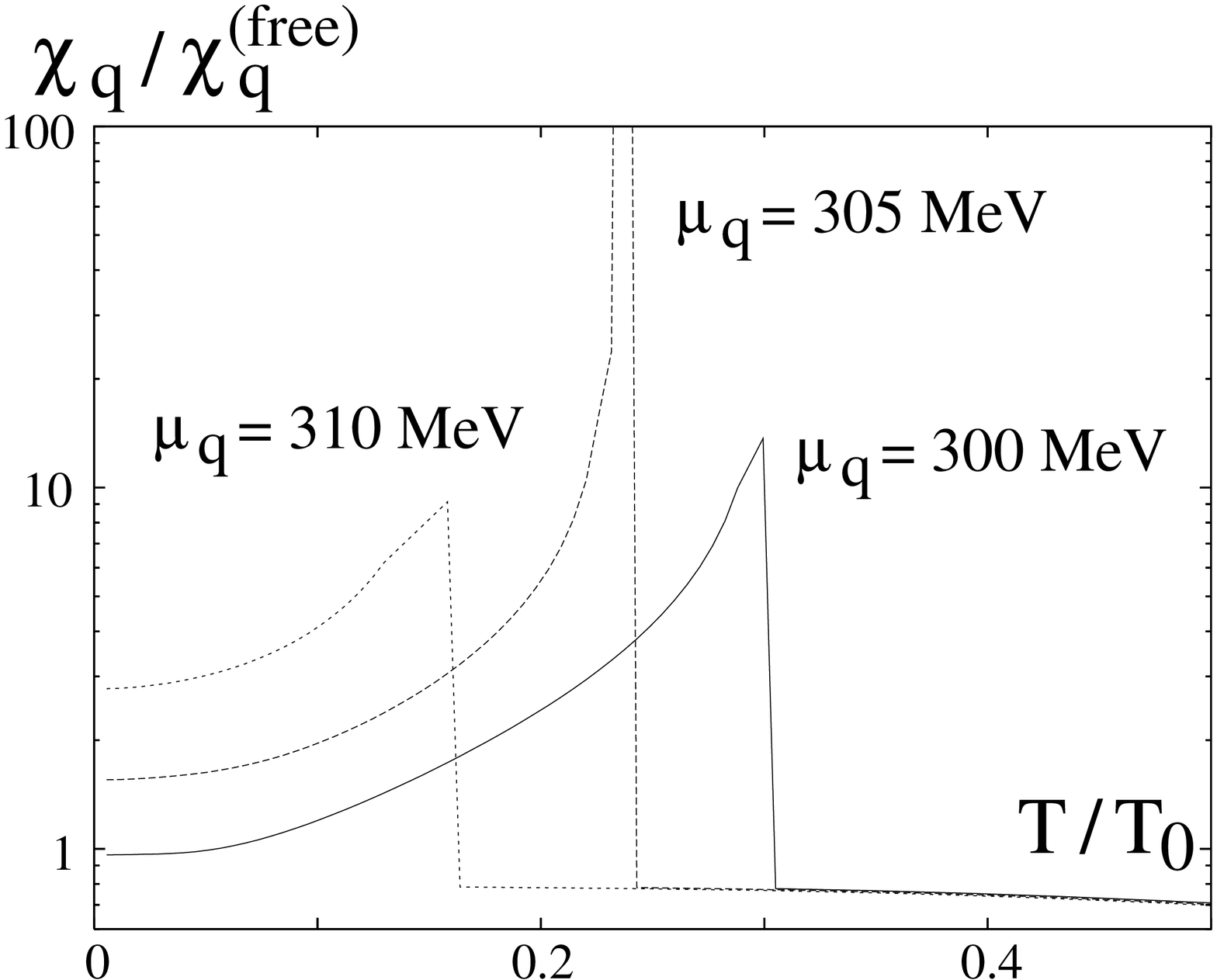}
\end{center}
\caption{\protect\label{fig:chiq} The quark number susceptibility $\chi_{q}$ for
different values of net--quark chemical potentials $\mu_q$ as a function of $T/T_0$ in
the chiral limit. The $\chi_q^{\rm (free)}$ is the quark number susceptibility for ideal
quark gas and $T_0 = 177$ MeV is the transition temperature at $\mu_q = \mu_I = 0$. The
calculations correspond to isospin symmetric system and the vector coupling constant
$G_V^{\rm (S)}=0.3\,G_S$. } \vspace*{0.5cm}
\end{figure}

Fig.~\ref{fig:phase} represents  the phase diagram of the NJL model related with the
chiral symmetry restoration in the $(T,\mu_q)$--plane obtained in the isospin symmetric
system and in the limit of vanishing current quark masses. The location of the
  boundary line that separates chirally broken from the symmetric
phase was found from the requirement of vanishing dynamical quark mass $M(T, \mu_q)=0$
when approaching from the side of the broken phase. Along the boundary line the order of
the chiral symmetry restoration transition is not unique. In the high quark density
regime  the phase transition is of the first order and terminates at the finite $T$ and
$\mu_q$ as the second order transition corresponding to the position of the tricritical
point (TCP). In the NJL model under the mean field approximation the location of the TCP
is determined by the condition of vanishing of the  second and the fourth order
coefficients in  the Taylor expansion of the thermodynamical potential:  $\Omega
(M,\mu_q,T)=a_0+a_2M^2+a_4M^4+O(M^6)$ applicable in the limit of $M\to 0$. For
temperatures above the TCP the transition stays of the second order as
expected in the Ginzburg--Landau theory.

The position of the boundary line and the TCP clearly depends on the values of the model
parameters \cite{Buballa,Asakawa:1989bq,Lutz:1992dv,Kitazawa:2002bc}. In
Fig.~\ref{fig:phase} the critical line was calculated for different strength of the
vector coupling $G_V$  at fixed values of $G_S$ and the momentum cut-off $\Lambda$. As
can be seen in Fig.~\ref{fig:phase}, an increase of $G_V$ results a decrease of
 $T_c$ and a shift  of $\mu_q^c$ towards larger values. This is to be
 expected \cite{Kunihiro} as the vector couplings $G_V^{(S)}$ and
 $G_V^{(V)}$ are related with repulsive interactions of
 constituent quarks.
The position of the TCP is also shifted to the lower temperature and higher $\mu_q$ with
increasing coupling  in the vector channel. A similar modification  of  position of the
phase diagram and the TCP is observed with  change of the scalar coupling $G_S$ and the
cut--off $\Lambda$. It is interesting to see that for a sufficiently large $G_V$ the TCP
disappears from the phase diagram and in the whole parameter range the phase transition
is of the second  order.

\subsubsection{Quark susceptibilities near the phase boundary}

With the thermodynamic potential and the self-consistent gap equations one can calculate
the net--quark number and isovector susceptibilities and study their sensitivity and
behavior near the phase transition.

To calculate $\chi_q$ and $\chi_I$ from Eq. (\ref{eq2.7}) we have to take into account
that the dynamical masses $M_f$ and the shifted chemical potentials $\tilde{\mu}_f$ are
implicitly dependent on $\mu_q$, $\mu_I$ and $T$. Consequently the susceptibilities
$\chi_{q,I}$ are controlled by derivatives of $M_f$ and $\tilde\mu_f$
\begin{eqnarray}
\chi_q &=& \frac{2N_c}{T}\sum_{f=u,d}\int\frac{d^3p}{(2\pi)^3}
    \Biggl[
  {}- \frac{M_f}{E_f}\frac{\partial M_f}{\partial\mu_q}
      \Bigl( n_f^{(+)}\bigl( 1 - n_f^{(+)} \bigr)
        {}-  n_f^{(-)}\bigl( 1 - n_f^{(-)} \bigr) \Bigr)
\nonumber\\
&&\qquad
     {}+ \frac{\partial\tilde{\mu}_f}{\partial\mu_q}
      \Bigl( n_f^{(+)}\bigl( 1 - n_f^{(+)} \bigr)
        {}+  n_f^{(-)}\bigl( 1 - n_f^{(-)} \bigr) \Bigr)
    \Biggr]\,,
\label{sus_q}\label{eq3.2}
\\
\chi_I &=& \frac{2N_c}{T}\int\frac{d^3p}{(2\pi)^3}
    \Biggl[
  {}- \frac{M_u}{E_u}\frac{\partial M_u}{\partial\mu_I}
      \Bigl( n_u^{(+)}\bigl( 1 - n_u^{(+)} \bigr)
        {}-  n_u^{(-)}\bigl( 1 - n_u^{(-)} \bigr) \Bigr)
\nonumber\\
&&\qquad
     {}+ \frac{\partial\tilde{\mu}_u}{\partial\mu_I}
      \Bigl( n_u^{(+)}\bigl( 1 - n_u^{(+)} \bigr)
        {}+  n_u^{(-)}\bigl( 1 - n_u^{(-)} \bigr) \Bigr)
   {}- (u \to d)
    \Biggr]\,,
\label{sus_I}\label{eq3.3}
\end{eqnarray}
where we have suppressed for simplicity  the $T$ and $\tilde{\mu}_f$ dependence of
$n_f^{(\pm)}$ distributions.

The NJL model does not exhibit confinement properties of QCD. Thus, there are  no
hadronic bound states and resonances in the chirally broken phase in the NJL  medium.
Instead, we are dealing with constituent quarks which can be viewed as quasi-particles
with the  temperature and density dependent mass. In the chirally symmetric phase
composition of the medium in the NJL model is not changed. At the chiral transition the
dynamical quark masses $M_f$ vanish and above $T_c$ the medium is populated by
interacting massless quarks. In addition, due to the momentum cut-off there is a
suppression of large momentum quark modes which is particularly efficient at high
temperature. The differences in the mass spectrum of the NJL model and the QCD as well as
suppression of the particle thermal phase-space will result in different quantitative
properties of quark number fluctuations. However, this does not exclude some possible
common features  of susceptibilities in QCD and in the NJL model related with the
restoration of the chiral symmetry.

Fig.~\ref{fig:chiq} shows the quark number susceptibility $\chi_q$ as a function of $T$
for different values of   $\mu_q$. The calculations correspond to the chiral limit and
were done for the fixed value of the vector coupling constant $G_V^{(S)}=0.3G_S$. The
net--quark fluctuations are normalized to that one for an ideal quark gas $\chi_q^{\rm
(free)} = N_c N_f (T^2/3 + \mu_q^2/\pi^2)$.

\begin{figure}
\begin{center}
\includegraphics[width=7.9cm]{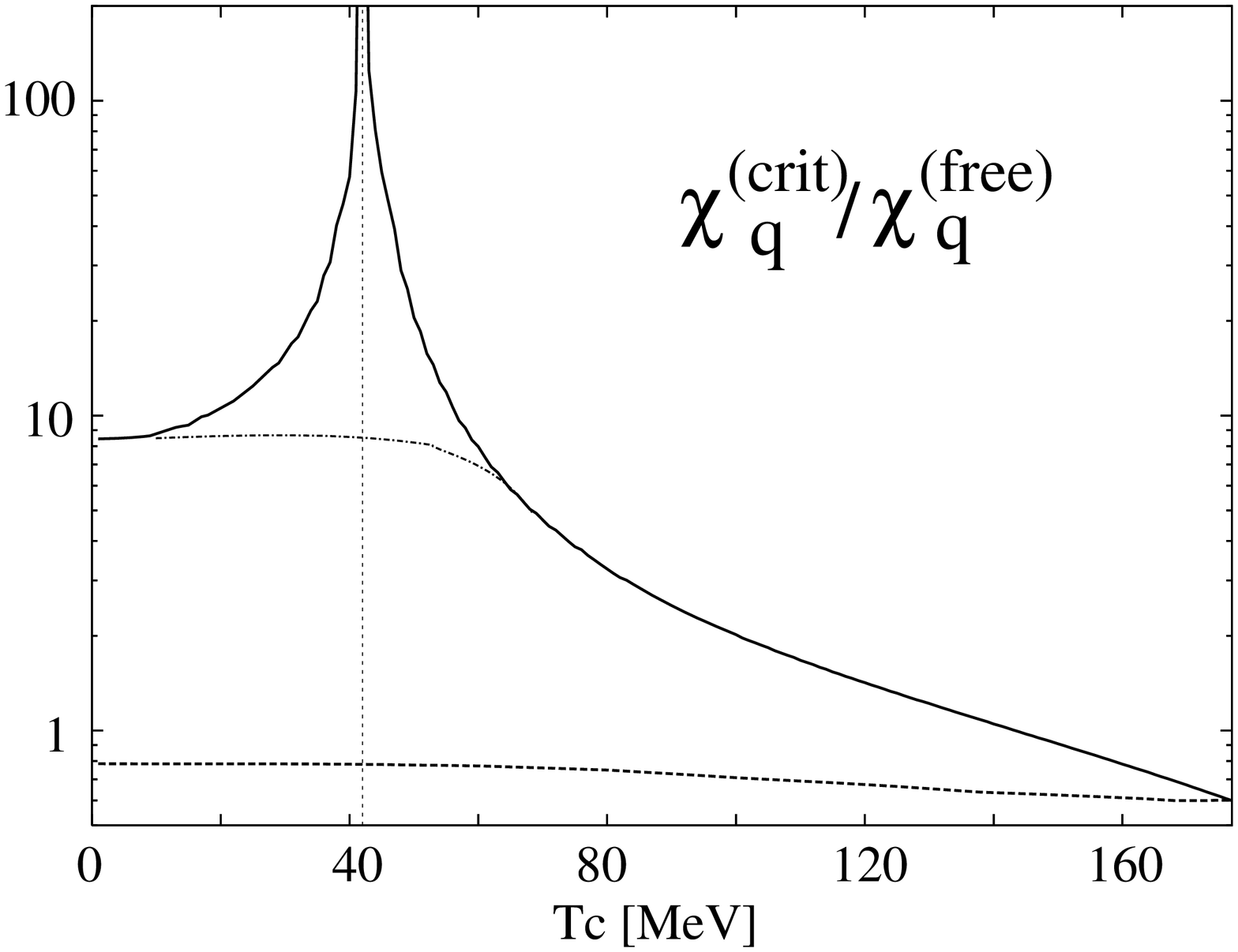}
\includegraphics[width=7.9cm]{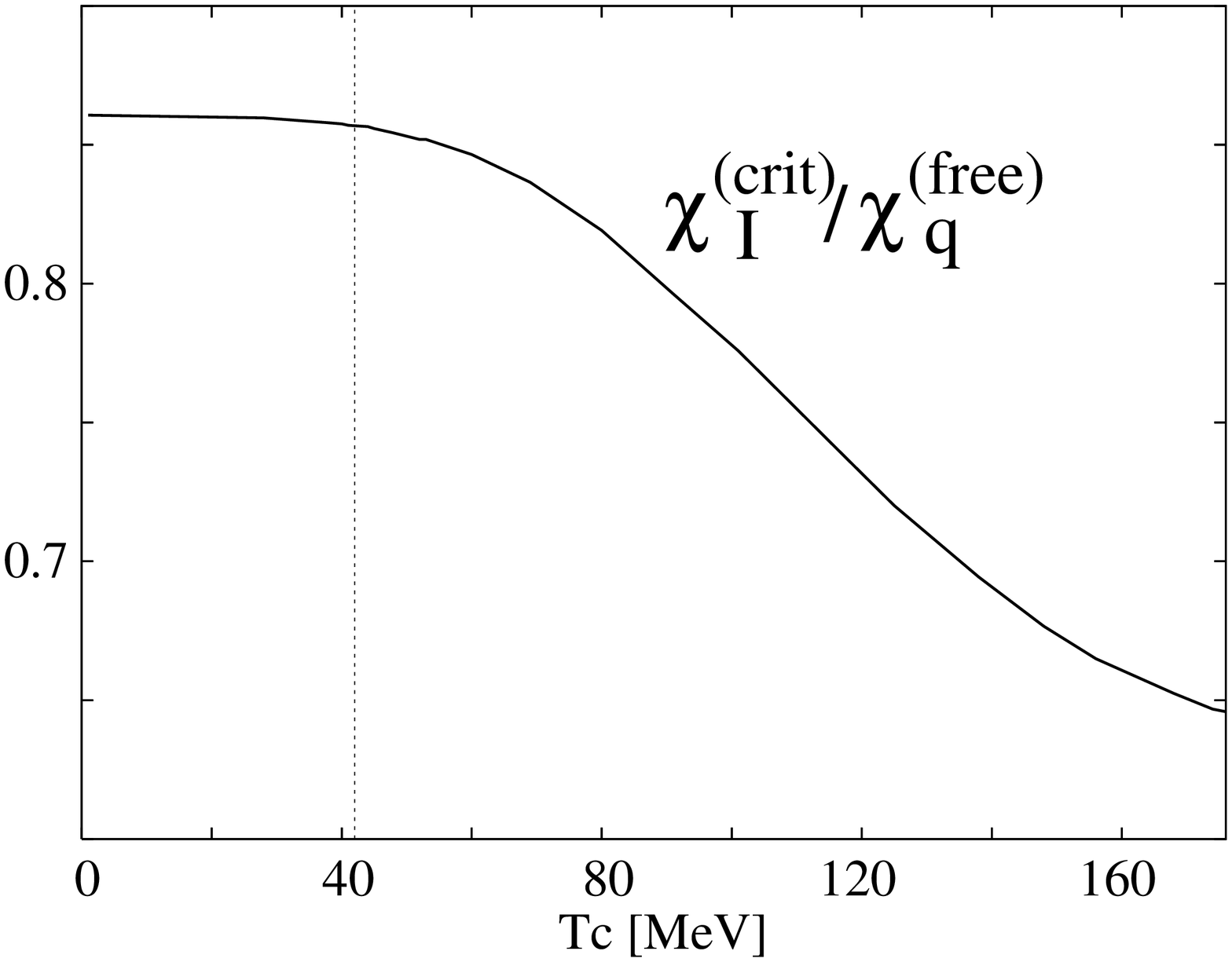}
\end{center}
\caption{\protect\label{tcp_q} The quark number (left) and isovector (right)
susceptibilities $\chi_{q,I}$ along the phase boundary line as a function of the
transition temperature $T_c$. In the left--hand  figure  the solid line denotes the
$\chi_q$ approaching from the chiral broken phase and the dashed line from the symmetric
phase. The vertical dotted-line indicates the position of the  tricritical point. The
calculations were done in the chiral limit in the  isospin symmetric system with  the
vector coupling constant $G_V^{\rm (S)}=0.3\,G_S$. } \vspace*{0.5cm}
\end{figure}
There are generic features in the  temperature dependence of $\chi_q$ for different
values of $\mu_q$. From  Fig.~\ref{fig:chiq} it is clear that the model  exhibits a phase
transition for all values of $\mu_q$. The transition temperature is strongly
$\mu_q$--dependent and decreases with increasing $\mu_q$. However, there is an essential
difference between the critical behavior  of $\chi_q$ at vanishing  and  at the finite
$\mu_q$.  For  $\mu_q\neq 0$ the susceptibility exhibits discontinuity at $T_c$ which
increases with  increasing  $\mu_q$. For $\mu_q=0$ such a discontinuous structure
disappears and instead at $T_c$ the susceptibilities shows a non-analytic structure which
results as discontinuity in higher moments of net--quark number fluctuations. Such a
properties of $ \chi_q$ are consistent with that expected if the phase transition is of
second order and belongs to universality class of three-dimensional $O(4)$ symmetric spin
models \cite{hatta,rg2}. To see it, let us construct an effective thermodynamic potential
in the near vicinity to the chiral transition. The relevant field is here a constituent
quark that carries  the dynamical  mass $M$. Performing the Taylor expansion of
$\Omega(M,T,\mu_q)$ as  power series around $M\simeq 0$ one gets:
\begin{equation}
\Omega^{}(T,\mu_q,M) \simeq  \Omega_0(T,\mu_q) {}+ \frac{1}{2}a(T,\mu_q)M^2 {}+
\frac{1}{4}b(T,\mu_q)M^4\,.\label{eqgl}
\end{equation}
Thus, $\Omega$ has a structure of  the Ginzburg--Landau (GL) potential where the
effective quark mass acts as  the sigma field. Following the GL theory and applying   the
mean field approximation  the  effective potential (\ref{eqgl}) is rewritten as
\begin{equation}
\Omega^{}(T,\mu_q;M_0) = \Omega_0(T,\mu_q)
{}-\frac{1}{4}\frac{a^2(T,\mu_q)}{b(T,\mu_q)}\,.\label{eqglm}
\end{equation}
where we have used the stationary condition $\partial \Omega/\partial M|_{M_0}=0$  and
introduced the stationary point
\\
$M_0 = \sqrt{-a/b}$. Approaching the critical line from the symmetric phase with $M=0$,
the quark number susceptibility  $\chi_q$ is obtained from Eq. (\ref{eqgl}) as
\begin{equation}
\chi_q^{\rm (sym)} = \frac{\partial^2\Omega_0}{\partial\mu_q^2}\,.\label{eqs}
\end{equation}

In the GL theory the second-order phase transition line (the $O(4)$ critical line) is
determined by the requirement that the coefficient   $a = 0$ and $b \neq 0$ in  Eq.
(\ref{eqgl}). In the near vicinity to   the critical point $(T_c,\mu_q^c)$ the
coefficient $a(T,\mu_q)$ can be parameterized in the MF approximation as
\begin{eqnarray}
a(T,\mu_q) \simeq  C (T - T_c) + D (\mu_q - \mu_{q}^c)\,,\label{eqa}
\end{eqnarray}
where  $C$ and $D$ are independent of $T$ and $\mu_q$.
Approaching the $O(4)$ critical line from the broken phase, the quark susceptibility is
calculated from Eqs. (\ref{eqglm}) and (\ref{eqa}) as:
\begin{equation}
\chi_q^{\rm (broken)} = \frac{\partial^2\Omega_0}{\partial\mu_q^2} {}-
\frac{D^2}{2b(T,\mu_q)}\,.\label{eqas}
\end{equation}
Comparing Eqs. (\ref{eqs}) and (\ref{eqas}) it is clear that  the second term in
(\ref{eqas}) gives just a  discontinuity of $\chi_q$ across the $O(4)$ critical line at
the finite $\mu_q$. While at $\mu_q=0$ the coefficient $D=0$ and $\chi^{\rm (
sym)}=\chi^{\rm (broken)}$ at $T=T_c$.

Considering the phase diagram in Fig.~\ref{fig:phase} we have already  indicated that for
large $\mu_q$ the NJL  model experiences the TCP. The right pannel of Fig.~\ref{fig:chiq}
shows the properties of the quark number susceptibility in the near vicinity and at TCP.
In the GL theory the position of the TCP is characterized  by the condition of  vanishing
$a(T,\mu_q)$ and $b(T,\mu_q)$ coefficients in the effective potential (\ref{eqgl}).
Consequently, from Eq. (\ref{eqas}) it is clear that the quark fluctuations $\chi_q$
should diverge at TCP. These expected properties of $\chi_q$ are clearly seen in
Fig.~\ref{fig:chiq}. With the present choice of parameters the  TCP is located at
$(T_c,\mu_q^c)=(42,305)$ MeV where $\chi_q\to \infty$.  Crossing TCP towards larger
$\mu_q$ the second order O(4) chiral phase transition is converted to the first order
where $\chi_q$ is finite and  has a gap at the critical temperature.

From  the perspective of heavy ion experiments  the properties of different
susceptibilities are of particular interest. This is because, fluctuations related with
the conserved charges are  experimentally   directly accessible . Since these are also
observables that are sensitive to the  critical behavior,  knowledge of the properties of
susceptibilities along the phase boundary line could give  insights how to verify the QCD
phase transition experimentally. Clearly, the quantitative structure of the phase diagram
and the position of the chiral endpoint is model dependent. Thus, also the position of
the QCD boundary line could be very different than that found  in
 the NJL model. However, the model  study  could still answer
a phenomenologically relevant question how to observe and how large is the critical
region along the phase transition where the fluctuations are sensitive to the singular
structure at the tricritical or chiral endpoint.

Fig.~\ref{tcp_q}  shows the net--quark susceptibility $\chi_q$ along the phase  boundary
line from Fig.~\ref{fig:phase}. The $\chi_q$ is quantified  as a function of the chiral
phase transition temperature $T_c$. The appearance of the TCP in the phase diagram
results in a non-monotonic behavior of $\chi_q$ along the transition line.  There is a
window of $\Delta T_c\simeq 30$ MeV above and $\Delta\mu_q^c \simeq 10$ MeV around TCP
where the fluctuations are sensitive to the appearance of the tricritical point. If the
TCP was absent in the phase diagram then the $\chi_q$ would be a monotonic function of
$T_c$ along the phase  boundary  as shown by  a dashed--dotted line in Fig.~\ref{tcp_q}.
Such a behavior is seen in Fig. 4--right in the isovector susceptibility  which is not
sensitive to the appearance of the TCP in the phase diagram \cite{probe1}.
 The  non-monotonic behavior of
$\chi_q$ is only seen from the side of the chirally broken phase. Approaching $T_c$ from
the chirally symmetric phase results in continuous  behavior of $\chi_q$ along the
boundary line. This is because  the  $\chi_q^{( \rm sym )}$ is finite at the chiral
transition in the whole parameter range as seen  in Fig.~\ref{tcp_q}. A difference
between $\chi_q^{(\rm sym )}$ and $\chi_q^{(\rm broken )}$ calculated along the phase
transition line measures the magnitude of discontinuity at the phase transition.
Fig.~\ref{tcp_q} shows that at $\mu_q^c=0$ this discontinuity vanishes and  at $T_c=0$
would be the largest if the phase diagram did  not experience the TCP.

In nucleus--nucleus collisions any    change of  the temperature and chemical potential
is correlated with the  corresponding  change of the c.m.s
 collision energy $\sqrt s$ \cite{our}. An increase of $\sqrt s$ results in
increasing  the temperature    $T$ and decreasing the quark chemical potential  $\mu_q$.
Thus, the critical region around tricritical or chiral endpoint $(\Delta \mu_q^c,\Delta
T_c)$ can be converted to that in the c.m.s. $\sqrt s$--energy in A--A collisions.
Admitting that the relation of $\mu_q^c$ and $T_c$ with $\sqrt s$ are the same as  for
chemical freezeout parameters extracted from experimental data \cite{our}  the $\Delta
T_c\simeq 30$ MeV window around the TCP would correspond to $\Delta s\sim 1$ A$\cdot$GeV.
Thus, within our crude estimates, we can expect that to observe any remnant of critical
fluctuations in A--A collisions one would need the $\sqrt s$ energy step to be within a
range of 1 AGeV.

\vskip 0.3cm  We  have discussed the properties of quark fluctuations in terms of
Nambu--Jona-Lasinio (NJL) model. The model was formulated at finite temperature and
chemical potentials related with the conservation of baryon number and isospin. Applying
the  mean field approach, we have shown how the fluctuations of different quark flavors
are changing across the phase boundary. Such a study is interesting from the perspective
of  heavy ion phenomenology and the lattice gauge theory. In the first case we have
indicated and quantified the non-monotonic structure of the  net--quark number
susceptibility along the phase boundary as the  method to identify the position of the
chiral endpoint. We have also discussed the critical region around this  point in the
context of heavy ion phenomenology.

\vskip 0.3cm We acknowledge interesting discussions with F. Karsch and J. Wambach. C.S.
also acknowledges fruitful   discussions with H. Fujii and B.~J.~Schaefer. K.R.
acknowledges stimulating discussions with S. Ejiri,  K.~Rajagopal, E.~V.~Shuryak  and
M.~A.~Stephanov. The work of B.F. and C.S. were supported in part by the Virtual
Institute of the Helmholtz Association under the grant No. VH-VI-041. K.R. acknowledges
partial support of the Gesellschaft f\"ur Schwerionenforschung (GSI)   and KBN under
grant 2P03 (06925).

\end{document}